\documentclass[prb,twocolumn,showpacs,amsmath,amssymb,superscriptaddress]{revtex4-1}
\usepackage{epsfig,epstopdf} % Transform eps to pdf
\usepackage{graphicx}% Include figure files
\usepackage{dcolumn}% Align table columns on decimal point
\usepackage{bm}
\usepackage{bbm}
\usepackage{ulem}
\usepackage{color}
\usepackage{xcolor}
\usepackage{slashed}
\usepackage{hyperref}
\usepackage{mathrsfs}
\usepackage{subfigure}
\usepackage{verbatim}
\usepackage{dsfont}
\usepackage{float}

\newcommand{\nc}{\newcommand}
\nc{\be}{\begin{equation}} \nc{\ee}{\end{equation}}
\nc{\bea}{\begin{eqnarray}} \nc{\eea}{\end{eqnarray}}
\nc{\bp}{\begin{pmatrix}} \nc{\ep}{\end{pmatrix}}
\nc{\ben}{\begin{enumerate}} \nc{\een}{\end{enumerate}}
\nc{\dg}{\dagger}
\nc{\ua}{\uparrow} \nc{\da}{\downarrow}
\nc{\rag}{\rangle} \nc{\lag}{\langle}

\begin{document}

\title{Topological invariants for three dimensional Dirac semimetals and four dimensional topological rotational insulators}

%% Notice placement of commas and superscripts and use of &
%% in the author list
\date{\today}

\author{Rui-Xing Zhang}
\affiliation{Department of Physics, The Pennsylvania State University, University Park, Pennsylvania 16802}
\author{Chao-Xing Liu}
\affiliation{Department of Physics, The Pennsylvania State University, University Park, Pennsylvania 16802}

\begin{abstract}
Dirac semimetal is a class of semi-metallic phase protected by certain types of crystalline symmetries, and its low-energy effective Hamiltonian is described by Dirac equations in three dimensions (3D). Despite of various theoretical studies, theories that describe the topological nature of Dirac semimetals have not been well established. In this work, we define a topological invariant for 3D Dirac semimetals by establishing a mapping between a 3D Dirac semimetal and a topological crystalline insulator in four dimension (4D). We demonstrate this scheme by constructing a tight-binding model for 4D topological crystalline insulators that are protected by rotational symmetry. A new type of topological invariant, "rotational Chern number", is shown to characterize the topology of this system. As a consequence of the rotational Chern number, gapless Dirac points are found on the 3D surface of this 4D system. For a slab with two surfaces, we find that the corresponding low-energy effective theory of two surface states can be directly mapped to that of a 3D Dirac semimetal, suggesting that topological nature of 3D Dirac semimetals can be characterized by rotational Chern number which is defined in 4D. Our scheme provides a new systematic approach to extract topological nature for topological semimetal phases.
\end{abstract}

\maketitle

%\newpage

\section{Introduction}
Topological semimetal (TSM) is a class of semi-metallic materials with topologically protected nodes \cite{wan2011,burkov2011a,burkov2011b,xu2011,wang2012,young2012}. The conduction and valence bands in a TSM usually touch at some momenta, forming nodal points or nodal lines. Examples of TSMs include three dimensional (3D) Weyl semimetals (WSMs) \cite{wan2011,burkov2011a}, 3D Dirac semimetals (DSMs) \cite{wang2012,wang2013,young2012} and nodal line semimetals \cite{burkov2011b}. 3D WSMs have an even number of band touching at discrete momentum points in the Brillouin zone (BZ), which requires the breaking of either time reversal symmetry (TRS) or spatial inversion symmetry. In the vicinity of the gapless nodes, the low-energy effective Hamiltonian is written as
\bea
H_{Weyl}= v_f {\bf k} \cdot {\bf \sigma},
\eea
which describes the physics of Weyl fermions in the high energy physics. Here ${\bf \sigma}=(\sigma_1,\sigma_2,\sigma_3)$ is the Pauli matrix for the pseudo-spin degree of freedom, and $v_f$ is the Fermi velocity, with its sign $sgn(v_f)=\frac{v_f}{|v_f|}$ determining the chirality of this Weyl node. Fermion doubling theorem (FDT) guarantees that the net chirality of all Weyl points in a WSM system is always zero. Theoretically, WSMs have been proposed in pyrochlore iridates \cite{wan2011}, topological insulator multilayer systems\cite{burkov2011a}, ferromagnetic HgCr$_2$Se$_4$\cite{xu2011} and noncentrosymmetric transition-metal monophosphides\cite{weng2015}. Experimentally, WSMs were demonstrated in TaAs systems \cite{lv2015,xu2015a,yang2015a} and very recently in MoTe$_2$ \cite{xu2016,jiang2016,liang2016,deng2016}. If two Weyl nodes with opposite chiralities meet in the BZ, they will form a four-fold degenerate Dirac node. Such Dirac point is usually unstable under perturbations, which results in the annihilation of two Weyl fermions and a finite energy gap. However, additional crystalline symmetries and TRS of the system will be possible to stabilize the Dirac point, giving rise to a 3D DSM. The simplest effective Hamiltonian of a DSM is a combination of two $H_{Weyl}$s with opposite chiralities,
\bea
H_{Dirac}=v_f\bp
{\bf k}\cdot\sigma & 0 \\
0 & -{\bf k}\cdot\sigma\\
\ep.
\label{Eq:Dirac}
\eea
Realizations of DSMs have been theoretically predicted \cite{wang2012,wang2013} and experimentally confirmed in Na$_3$Bi \cite{liu2014a,xu2015b} and Cd$_3$As$_2$ systems \cite{neupane2014,liu2014b,borisenko2014}. In addition, TSMs with nodal lines have also been theoretically proposed for graphene networks \cite{weng2015} and Cu$_3$PdN\cite{yu2015}. Exotic physical phenomena have been theoretically proposed as a consequence of topologically non-trivial band structures, including TSM induced axionic insulating phases \cite{wan2011,wang2013b,zhang2016} and chiral anomaly induced negative magneto-resistance \cite{son2012,zyuzin2012,aji2012}, etc. Some of them, such as negative magneto-resistance, have been observed in TSMs experimentally\cite{xiong2015,huang2015,liang2015}.

TSM phases are known to be closely related to topological insulating phases. We may take a WSM as an example and consider a sphere in the BZ that encloses one Weyl point. A gap exists and separates the conduction bands from the valence bands on this sphere. As a result, one can define a topological invariant, the first Chern number $C^{(1)}$, on this 2D sphere, and the non-zero value of $C^{(1)}$ implies the existence of the Weyl point. One can show that $C^{(1)}$ also defines the chirality of this Weyl point. A direct physical consequence of the nontrivial $C^{(1)}$ is the existence of surface Fermi arc states \cite{wan2011}. On the other hand, the first Chern number is also used to characterize the quantum Hall state \cite{thouless1982} and the quantum anomalous Hall state \cite{yu2010,chang2013} in 2D. Therefore, one can see the close relationship between the 3D WSMs and the 2D quantum Hall states.

As for DSMs, debates arise about whether DSMs are topological or not, due to the lack of a well-defined topological invariant. A straightforward approach is to identify a possible topological invariant of DSMs in terms of chirality, similar to what we did for a WSM. However, since the net chirality of a Dirac point is always zero, it is possible to introduce additional terms to the effective Hamiltonian in Eq. [\ref{Eq:Dirac}], such as the cubic terms in the effective model of Na$_3$Bi\cite{wang2012} to couple the two Weyl fermions while preserving the gapless nature. As a result, the chirality is not well-defined when two Weyl points are coupled. Another possible scheme assumes that the existence of double Fermi arc states (DFAS) on the surface of DSMs can reveal the topological nature of a bulk Dirac point \cite{yang2014}. However, a recent work \cite{kargarian2015} shows that these DFAS can be detached from the bulk Dirac point by adding symmetry-allowed perturbations. Therefore, the DFAS in both Na$_3$Bi and Cd$_3$As$_2$ systems can not characterize the stability of the bulk Dirac point. %In other words, we can gap out the bulk Dirac point in these systems with a rotation symmetry breaking term. Then the system is driven into a TRS invariant topological insulator, while gapless surface states (DFAS) still exists.}

A well-defined topological invariant of DSMs must emphasize the important role of crystalline symmetries. The above topological invariant schemes, however, overlook the protection mechanism of crystalline symmetry, and that is why the corresponding topological invariants are not "stable". In another recent work\cite{yang2015b}, it is proposed that topological charges of DSMs can be defined as topological invariants of zero-dimensional subsystems defined on the rotation axis.

%Researchers also applies this idea to the Dirac semi-metal phase, which possesses two Weyl fermions with opposite chiralities at the same momentum in the BZ, in 3Ds and search for the corresponding topological insulating phases, which share the same topological invariant as Dirac semi-metals, in 2Ds. In contrast to Weyl semi-metals, the stability of Dirac semi-metals requires additional symmetry protection, including crystalline symmetry or the combination of crystalline symmetry and time reversal symmetry [XXX]. However, crystalline symmetry in 3Ds is different from that in 2Ds. For example, screw axis can exist in 3Ds but not in 2Ds. Therefore, despite of some efforts in some special cases, a general and unified definition of topological invariants for Dirac semi-metals, as well as other topological nodal phases that are protected by crystalline symmetry, has not been achieved. ({\bf we probably need to be careful about this statement}).

In this paper, we consider an alternative approach to define topological invariants for 3D DSMs. Our motivation is inspired by another perspective of WSMs: a single Weyl fermion can exist at the boundary of a four dimensional (4D) generalization of the quantum Hall insulator (QHI) state with a non-trivial second Chern number $C^{(2)}$ \cite{qi2008,golterman1993,creutz2001}. Consequently, a WSM with two Weyl fermions (the minimal number of Weyl fermions in 3D bulk systems) can be viewed as a slab of a 4D quantum Hall system with an auxiliary fourth dimension. The slab direction is chosen to be along the auxiliary fourth direction and the low energy effective physics, which includes two gapless Weyl fermions at two opposite 3D surfaces of a slab, recovers that of a WSM. According to this approach, the second Chern number, defined in the bulk of a 4D QHI, also guarantees the occurrence of 3D gapless Weyl fermions at the boundary. Therefore, we can regard the second Chern number $C^{(2)}$ in 4D as an alternative topological invariant for WSMs in 3D. Another advantage is that anomalous topological response of a WSM, namely quantum anomaly\cite{zyuzin2012}, can be easily understood in this approach. For example, the chiral anomaly of 3D Weyl fermions can be understood as the charge (as well as chirality) non-conservation on one surface of a 4D QHI as a result of a transverse current flow along the auxiliary fourth direction, when an electric field is applied in parallel to a magnetic field in the other 3D space\cite{qi2008}. This effect is known as the Callan-Harvey effect\cite{callan1985}, and in 2D, a similar charge non-conservation takes place on the 1D edge of a 2D quantum Hall state.

In this work, we will generalize this idea and search for a 4D topological crystalline-symmetry-protected insulating (TCI) phase\cite{fu2011}, whose low-energy effective theory in a slab configuration can reproduce that of a 3D DSM. Generally, we can consider a system in 4D and neglect any symmetry operation that is related to the auxiliary fourth dimension. As a result, the corresponding symmetry group for the 4D slab recovers a 3D space group, and provides necessary symmetry operations for a 3D DSM. %Throughout this paper, we will limit our discussions to point group symmetries. A natural generalization of our theory to space group symmetries (such as screw rotation and glide reflection symmetries) will be left future works.}

The paper is organized as follows. As a warm up, we briefly review the basic theory of 4D QHI in section \ref{sec:II}. By constructing a 3D WSM from a 4D QHI, we demonstrate our basic scheme of constructing a 3D bulk semi-metallic phase from a 4D topological insulator slab. In section \ref{sec:III}, we construct the central model in our work and show that it is a 4D TCI phase protected by four-fold rotation symmetry, thus dubbed "topological rotational insulator" (TRI). The topology of this model is characterized by a new topological invariant "rotational Chern number" (RCN), which is only well defined in 4D systems. The RCN guarantees the existence of representation-dependent chiral surface modes, where a surface Weyl fermion forms when two chiral surface modes cross. In section \ref{sec:IV}, we show that when additional symmetries are considered, two surface Weyl fermions with opposite chiralities are forced to appear at the same momentum position, forming a surface Dirac point (SDP). We identify two inequivalent minimal symmetry requirements to stabilize the SDPs. In section \ref{sec:V}, we finally establish the mapping between our 4D TRI model and 3D DSM systems by projecting the Hamiltonian onto the surface states bases. This equivalence between a 4D TRI system and a 3D DSM system allows us to utilize RCN as the topological invariant for a rotation-symmetry-protected DSM. In section \ref{sec:VI}, the conclusion and discussion are presented. In the appendix, we show in details the construction of symmetry operations in our model, with the help of Clifford algebra, as well as a detailed calculation of the surface state projection.

%In the previous section, we have shown that a TRS invariant (4+1)D QHI can be mapped to a (3+1)D TRS broken WSM with the help of a thin film dimension reduction construction scheme. It is natural to ask whether a (3+1)D Dirac semimetal (DSM) can be mapped to a (4+1)D surface Dirac point protected by certain crystalline symmetries. In this section, we give an affirmative answer to this question by explicitly construct a (4+1)D topological crystalline insulator (TCI) model. We show that this TCI phase is protected by in-plane four-fold rotation symmetry, and characterize its bulk topology with a new topological invariant: rotation Chern number (RCN). The minimal symmetry requirements to stabilize a surface Dirac point is carefully studied, leading to two different protection mechanisms. From this, we map this (4+1)D TCI model in a thin film configuration to the realistic DSMs.

\section{4D quantum Hall insulators and 3D Weyl semimetals}\label{sec:II}
%\begin{figure*}[t]
%\begin{minipage}{0.69\textwidth}
%  \centering
%    \includegraphics[width=\textwidth]{4Dweyl.eps}
%  \end{minipage}\hfill
%  \begin{minipage}{0.28\textwidth}
%  \centering
%  \caption{}
%\end{minipage}
%\label{Fig:4Dweyl}
%\end{figure*}
\begin{figure*}[t]
\centering
\includegraphics[width=\textwidth]{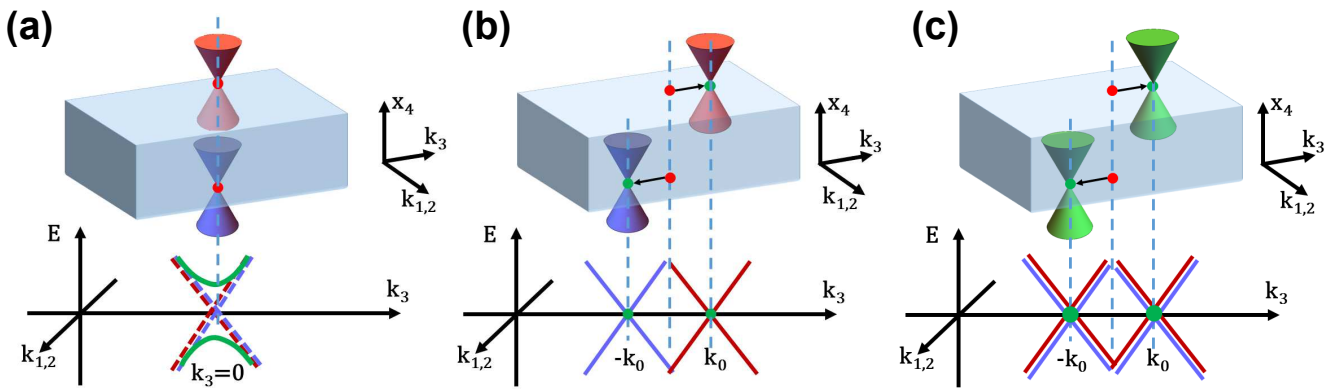}
\caption{In (a), brown (purple) lines cross to form a surface Weyl fermion with $+1$ ($-1$) chirality. These two surface Weyl fermions are projected together in the 4D QHI thin film and hybridization effects can induce a bulk gap (green lines). In (b), surface Weyl fermions are shifted in opposite way, and this momentum separation stabilize these Weyl points, forming a 3D WSM in the thin film system. In (c), we consider a 4D TRI system with two SDPs projected together. Each SDP consists of two surface Weyl fermions with opposite chiralities. SDPs are shifted to $\pm k_0$ to stabilize their gapless nature, forming a 3D DSM.}
\label{Fig:4Dweyl}
\end{figure*}
To warm up, let us start by reviewing the well-known 4D Dirac model which describes a 4D QHI\cite{qi2008,golterman1993,creutz2001}. A lattice version of the 4D Dirac model is given by
\bea
H^{4D}_{Dirac}=\sum_{i=1}^4 \sin k_i \gamma_i +m(k)\gamma_5,
\label{Eq:4D Dirac}
\eea
where $m(k)=m_0+B_0\sum_{i=1}^4\cos k_i$. The five fundamental $4\times4$ $\gamma$ matrices are defined as
\bea
\gamma_1&=&\sigma_1\otimes\tau_3,\ \gamma_2=\sigma_2\otimes\tau_3 \nonumber \\
\gamma_3&=&\sigma_0\otimes\tau_1,\ \gamma_4=\sigma_0\otimes\tau_2 \nonumber \\
\gamma_5&=&\sigma_3\otimes\tau_3
\eea
with the Pauli matrices $\sigma_i$ and $\tau_i$ ($i=1,2,3$) characterizing different degrees of freedom. $\sigma_0$ and $\tau_0$ denote the $2\times 2$ identity matrices. The $\gamma$ matrices satisfy the anti-commutation relation, $\{\gamma_i,\gamma_j\}=0$ for $i\neq j$. The topology of the Hamiltonian (Eq. [\ref{Eq:4D Dirac}]) is characterized by the second Chern number $C^{(2)}$,
\bea
C_2=\frac{3}{8\pi^2}\int d^4k \epsilon^{abcde}\hat{d}_a \partial_{k_1} \hat{d}_b \partial_{k_2} \hat{d}_c \partial_{k_3} \hat{d}_d \partial_{k_4} \hat{d}_e,
\eea
where we have defined
\bea
{\bf d}=(\sin k_1,\sin k_2, \sin k_3, \sin k_4, m(k))
\eea
and $\hat{d}={\bf d}/|{\bf d}|$. Any change in $C^{(2)}$ can be only achieved via a topological phase transition, which is accompanied by an energy gap closing of energy bands. To map out the phase diagram, the energy spectrum can be solved analytically as
\bea
E(k)=\pm\sqrt{\sum_{i}\sin^2 k_i+m(k)^2}.
\eea
By solving $E(k_i,m_0,B)=0$, we arrive at the following topological phases that are characterized by different second Chern number
\begin{equation}
C^{(2)}=
\left\{
\begin{aligned}
&0,& \ m_0<-4B_0 \\
&1,& \ -4B_0<m_0<-2B_0 \\
&-3,&\ -2B_0<m_0<0 \\
&3,&\  0<m_0<2B_0 \\
&-1,&\ 2B_0<m_0<4B_0 \\
&0,&\  m_0>4B_0
\end{aligned}
\right..
\label{Eq:2nd Chern number}
\end{equation}
According to the bulk-boundary correspondence, a non-zero second Chern number $C^{(2)}$ in 4D gives rise to $|C^{(2)}|$ copies of Weyl fermions living on the 3D surfaces. An odd number of surface Weyl fermions are anomalous for a single surface, since in any realistic 3D systems, only an even number of Weyl fermions can occur according to the FDT. However, if we consider the number of Weyl fermions on both surfaces, the chiralities of Weyl fermions at opposite surfaces cancel each other, thus recovering the FDT.

%({\bf could you please integrate the next paragraph to the introduction part? })
%On the other hand, Weyl fermions are found to naturally exist in (3+1)D solid systems which are known as "Weyl semimetals" (WSM). In a WSM, two non-degenerate band touches each other to form a Weyl point, whose low-energy effective model exactly reproduces that of a Weyl fermion. These Weyl points are characterized by a topological charge, which is the chirality of a Weyl point. For every Fermi surface that encloses a Weyl point, the topological charge is the first Chern number $C^{(1)}$ on this (2+1)D Fermi surface. Also, the net chirality of Weyl points in a WSM is forced to vanish, to avoid the breaking of FDT. For example, in a time reversal symmetry (TRS) breaking WSM, the minimal number of Weyl points is two: one Weyl point carries a topological charge $+1$, while the other carries a $-1$ charge.

Next we will establish a mapping between a 4D QHI and a 3D WSM. We choose a slab configuration along the $x_4$ direction and periodic boundary conditions for the $x_i$ ($i=1,2,3$) direction, so that $k_{i}$ is still a good quantum number. In a "thin film" limit, surface Weyl fermions on both surfaces can be projected together, as shown in Fig. [\ref{Fig:4Dweyl}] (a). This forms a four-fold degeneracy (dashed lines) and this 4D QHI thin film can now be viewed as a 3D bulk system. This degeneracy is generally unstable under perturbations (solid green lines in Fig. [\ref{Fig:4Dweyl}] (a)). To stabilize this gapless point, we can shift the Weyl points from different surfaces in opposite momentum directions by replacing $k_3$ with $k_3+k_0(\frac{2x_4}{L}-1)$ in the slab configuration ($0\leq x_4 \leq L$). As shown in Fig. [\ref{Fig:4Dweyl}] (b), due to momentum separation, the stability of 3D Weyl points is protected by translation symmetry. The low-energy theory of this 4D QHI thin film is now equivalent to a 3D WSM with two bulk Weyl points at $\pm k_0$. This provides us with a new dimension reduction scheme to construct a 3D bulk topological semi-metallic phase from a 4D bulk topological insulating phase.

\section{A simple model of a 4D topological rotational insulator} \label{sec:III}
In this section, we will construct a theory to describe a 4D TRI phase protected by the four-fold rotation symmetry in the $x_1$-$x_2$ plane. The Hamiltonian of this 4D TRI model is written as %To start with, the (4+1)D QHI model provide us with a good example of a (4+1)D system with non-trivial topology. Therefore, we hope to use $H^{4D}_{Dirac}$ as a building block to construct our TCI theory. In addition, $C^{(2)}$ needs to be trivialized in our new system, while leaving the topology non-trivial. Then a simple model that one can immediately come up with should look like this:
%\be
%H^{4D}_0=
%\bp
%H^{4D}_{Dirac}(C^{(2)}=1) & 0 \\
%0 & H^{4D}_{Dirac}(C^{(2)}=-1)
%\ep
%\ee
%Here $H^{4D}_{Dirac}(C^{(2)}=\pm1)$ represents a 4D Dirac model with a second Chern number $C^{(2)}=\pm1$. We make use of the fact that second Chern number is defined only for occupied bands, and the $C^{(2)}$s for both the electron part and hole part add up to zero. Then $-H^{4D}_{Dirac}(C^{(2)}=1)$ has $C^{(2)}=-1$ and a simple form of $H^{4D}_0$ is
\be
H^{4D}_0=
\bp
H^{4D}_{Dirac} & 0 \\
0 & -H^{4D}_{Dirac}
\ep.
\label{Eq:H4D_general}
\ee
with the basis functions given by
\be
|\Psi\rangle=(|\frac{1}{2}\rag_1,|\frac{1}{2}\rag_2,|\frac{3}{2}\rag_1,|\frac{3}{2}\rag_2,
|-\frac{3}{2}\rag_1,|-\frac{3}{2}\rag_2,|-\frac{1}{2}\rag_1,|-\frac{1}{2}\rag_2)^T
\label{Eq:bases}
\ee
Here $|\frac{J}{2}\rag_i$ represents a basis with total angular momentum $\frac{J}{2}$ and an additional index $i=1,2$. We have assumed that the system has a four-fold rotation symmetry $C_4$ in the $x_1$-$x_2$ plane, which is defined as $C_4{\bf x}=(x_2,-x_1,x_3,x_4)$ with ${\bf x}=(x_1,x_2,x_3,x_4)$. Based on the form of $H^{4D}_0$, we extend the previous five $4\times 4$ $\gamma$ matrices to the following seven $8\times 8$ $\Gamma$ matrices,
\bea
\Gamma_1&=&s_3\otimes\sigma_1\otimes\tau_3,\ \Gamma_2=s_3\otimes\sigma_2\otimes\tau_3, \nonumber \\
\Gamma_3&=&s_3\otimes\sigma_0\otimes\tau_1,\ \Gamma_4=s_3\otimes\sigma_0\otimes\tau_2, \nonumber \\
\Gamma_5&=&s_3\otimes\sigma_3\otimes\tau_3,\ \Gamma_6=s_1\otimes\sigma_0\otimes\tau_0, \nonumber \\
\Gamma_7&=&s_2\otimes\sigma_0\otimes\tau_0,
\label{Eq:8by8 Gamma}
\eea
where anti-commutation relations $\{\Gamma_i,\Gamma_j\}=0$ for ($i\neq j$) are satisfied. All other $8\times 8$ $\Gamma$ matrices can be generated as
\bea
\Gamma_{ij}&=&\frac{1}{i}\Gamma_i\Gamma_j\ (i\neq j) \nonumber \\
\Gamma_{ijk}&=&\frac{1}{i}\Gamma_i\Gamma_j\Gamma_k\ (i\neq j\neq k).
\eea
With the $\Gamma$ matrices, $H^{4D}_0$ can be written in a compact form
\bea
H^{4D}_0=\sum_{i=1}^4 v_f k_i\Gamma_i+m(k)\Gamma_5
\label{Eq:H4D}
\eea
in the continuum limit with $m(k)=(m_0+4B_0)-\frac{B_0}{2}\sum_{i=1}^4 k_i^2$. Here Fermi velocity $v_f$ is assumed to be positive.

\begin{figure}[t]
\centering
\includegraphics[width=0.5\textwidth]{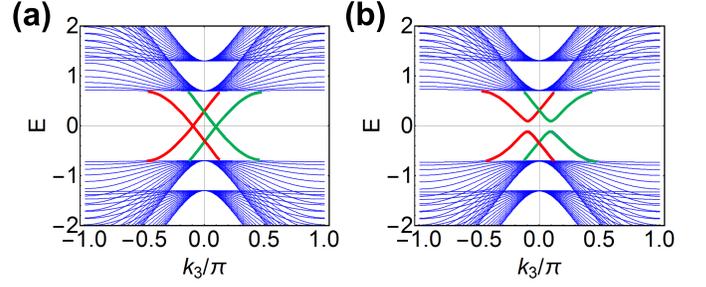}
\caption{Surface spectrums are calculated in a slab geometry stacking along $x_4$ direction for $(k_1,k_2)=(0,0)$ in (a) and $(k_1,k_2)=(0.1,0)$ in (b). Green (red) lines cross to form a SDP on the top (bottom) surface. We have chosen the parameters to be $m_0=-3,B_0=1,k_0=0.1\pi,v_f=1$.}
\label{Fig:4D dispersion}
\end{figure}

Let us consider a slab configuration along the $x_4$ direction ($0\leq x_4\leq L$) to study surface spectrum. We take $m_0=-3B$ so that $\pm H^{4D}_{Dirac}$ carries $C^{(2)}=\pm 1$. At either top or bottom surface, two surface Weyl fermions with opposite chiralities will show up at $k_1=k_2=k_3=0$ to form a surface Dirac point (SDP). As we have discussed in a 4D QHI system, we can redefine $k_3\rightarrow k_3+k_0(\frac{2x_4}{L}-1)$ to shift the SDP on the top (bottom) surface to $k_3=k_0(-k_0)$. Alternatively, we can introduce a shift term
\be
H_{shift}=k_0s_3\otimes\sigma_3\otimes\tau_0
\ee
which shifts the SDPs in an equivalent way. The expected surface spectrum of $H^{4D}_{0}$ is shown in Fig. [\ref{Fig:4Dweyl}] (c), where brown (purple) lines represent a Weyl fermion with its chirality being $+1(-1)$. Numerically, we verified our expectation by plotting the surface spectrum in Fig. [\ref{Fig:4D dispersion}] (a) with $k_1=k_2=0$. As is shown, the gapless surface state from top surface (green lines) appear at $k_3=k_0$ and it is numerically confirmed to be a four-fold degenerate SDP. At $k_3=-k_0$, we find the SDP from the bottom surface (red lines). Fig. [\ref{Fig:4D dispersion}] (b) shows the surface spectrum at $(k_1,k_2)=(0.1,0)$, and surface states are found to be gapped. This confirms that SDPs only appear at the $k_3$ axis.

The existence of gapless SDPs indicates the non-trivial bulk topology of $H^{4D}_0$. To understand its topological nature, we first emphasize the role of in-plane four-fold rotation symmetry $C_4$. Under the bases Eq. [\ref{Eq:bases}], we write down the representation of $C_4$ operator as
\bea
C_4=e^{iJ_{12}\frac{\pi}{2}}
\eea
with the rotation symmetry generator
\bea
J_{12}&=&
\bp
\frac{1}{2}\tau_0 & 0 & 0 & 0 \\
0 & \frac{3}{2}\tau_0 & 0 & 0 \\
0 & 0 & -\frac{3}{2}\tau_0 & 0 \\
0 & 0 & 0 & -\frac{1}{2}\tau_0 \\
\ep.
\eea
In general, we are able to generate an in-plane rotation operation $R_{\theta}$ for an arbitrary rotation angle $\theta$ as $R_{\theta}=e^{iJ_{12}\theta}$. In the continuum limit, $H^{4D}_0$ possesses the full in-plane rotation symmetry and is invariant under $R_{\theta}$. However, placing $H^{4D}_0$ on a lattice reduces $R_{\theta}$ to a discrete four-fold rotation symmetry $C_4=R_{\frac{\pi}{2}}$. Specifically, one can show that
\bea
C_4H^{4D}_0({\bf k})C_4^{-1}=H^{4D}_0(-k_2,k_1,k_3,k_4)
\eea
where ${\bf k}=(k_1,k_2,k_3,k_4)$.

Now we are ready to analyze the bulk topology of this model. We notice that the entire $k_3$-$k_4$ plane ($k_1=k_2=0$) is invariant under $C_4$ rotation, which is called a rotation invariant plane (RIP). On this RIP, the full Hamiltonian can be written in a block diagonal form as
\bea
H^{4D}_0(0,0,k_3,k_4)
=
\bp
h_{\frac{1}{2}} & 0 & 0 & 0 \\
0 & h_{\frac{3}{2}} & 0 & 0 \\
0 & 0 & h_{-\frac{3}{2}} & 0  \\
0 & 0 & 0 & h_{-\frac{1}{2}} \\
\ep
\eea
where each block $h_{\frac{J}{2}}$ belongs to one irreducible representation of $C_4$. Here $\frac{J}{2}$ is the angular momentum of the corresponding representation, and each $h_{\frac{J}{2}}$ is a $2\times2$ matrix,
\be
h_{\pm\frac{1}{2}}=-h_{\mp\frac{3}{2}}=
\bp
m(k_3,k_4) & \pm v_f(k_3-i k_4) \\
\pm v_f(k_3+i k_4) & -m(k_3,k_4) \\
\ep
\label{Eq:Block of H4D0}
\ee
with $m(k_3,k_4)=(m_0+4B_0)-\frac{B_0}{2}(k_3^2+k_4^2)$. We recognize that each block $h_J$ is equivalent to a 2D quantum anomalous Hall model\cite{qi2008}. Therefore, for each representation, we can define a corresponding first Chern number $C^{(1)}_J$ which explains the nontrivial gapless states on the surface:
\be
C^{(1)}_{\frac{1}{2}}=1,\ C^{(1)}_{-\frac{1}{2}}=1,\ C^{(1)}_{\frac{3}{2}}=-1,\ C^{(1)}_{-\frac{3}{2}}=-1.
\label{Eq:Chern number set}
\ee
The above set of $C^{(1)}$ is dubbed "rotational Chern number" (RCN), in analogous to the definition of mirror Chern number (MCN) in a 3D TCI\cite{hsieh2012}. We would like to emphasize that RCN is only well defined in a d-dimensional system with $d\geq4$, where a RIP can be found in the bulk Brillouin zone. As long as $C_4$ rotation symmetry is preserved, one can never couple these chiral edge modes from different representations and RCN is always well defined. Therefore, the gapless nature of nontrivial surface states is protected by in-plane rotation symmetry $C_4$. Together with the definition of RCN, we identify this model as a 4D rotation-symmetry-protected TCI, dubbed "topological rotational insulator" (TRI).

Because of the block-diagonal form of $H^{4D}_0$ (Eq. [\ref{Eq:H4D_general}]), one might wonder whether it is possible to define an equivalent topological invariant by counting the difference between second Chern number $C^{(2)}$ in each block. We notice that this definition relies on the special form of $H^{4D}_0$, which was initially assumed for simplicity. In the next section, we will introduce symmetry-allowed perturbation terms that explicitly destroy the block-diagonal form of $H^{4D}_0$. With these terms, $C^{(2)}$ is not well-defined, but the gapless surface states still survive as a result of RCNs.

\section{Additional essential symmetries for a stable surface Dirac point} \label{sec:IV}
In the previous section, we have successfully constructed a 4D TRI model and show that the topological invariant of this model can be described by RCN based on irreducible representation of rotation symmetry, which leads to non-trivial surface states on the boundary. In Fig. [\ref{Fig:4D dispersion}] (a), we found that there is a four-fold degenerate SDP on each surface. However, the RCN only guarantees the gapless nature of surface states, but says nothing about degeneracy. Therefore, one might wonder whether additional symmetries are required to protect this four-fold degeneracy.

Let us first identify other important symmetries of $H^{4D}_0$. We find that the TRI Hamiltonian is also invariant under the following crystalline symmetry operations,
\bea
\text{Spatial inversion}:\ {\cal I}{\bf x}&=&-{\bf x} \nonumber \\
\text{In-plane Mirror}:\ M_1{\bf x}&=&(-x_1,x_2,x_3,x_4).
\eea
Based on our bases choice and Clifford algebra, the matrix representation of the symmetry operators are
\bea
{\cal I}&=&\Gamma_5=s_3\otimes\sigma_3\otimes\tau_3 \nonumber \\
M_1&=&\Gamma_1\Gamma_6=is_2\otimes\sigma_1\otimes\tau_3.
\eea
To be specific, we find that
\bea
{\cal I}H^{4D}_0({\bf k}){\cal I}^{-1}&=&H^{4D}_0(-{\bf k}) \nonumber \\
M_1H^{4D}_0({\bf k})M_1^{-1}&=&H^{4D}_0(-k_1,k_2,k_3,k_4).
\eea
What is more, $H^{4D}_0$ is also found to be invariant under an anti-unitary symmetry operation $\Theta$, with
\bea
\Theta=\Gamma_2\Gamma_5\Gamma_7K=is_2\otimes\sigma_1\otimes\tau_0K,
\eea
where $K$ is the complex conjugate operation. Notice that $\Theta$ flips the sign of $k_{1,2,3}$ when operating on $H^{4D}_{0}$,
\bea
\Theta H^{4D}_{0}(k_{1,2,3},k_4)\Theta^{-1}=H^{4D}_{0}(-k_{1,2,3},k_4).
\eea
In addition,
\bea
\Theta H^{4D}_{tf}(k_{1,2,3},x_4)\Theta^{-1}=H^{4D}_{tf}(-k_{1,2,3},-x_4),
\eea
where $H^{4D}_{tf}$ is the form of $H^{4D}_0$ in a thin film configuration with the open boundary condition along the $x_4$ direction. Therefore, $\Theta$ transforms a SDP at $k_3=k_0$ on the top surface to another SDP at $k_3=-k_0$ on the bottom surface. We emphasize that $\Theta$ is NOT the time reversal symmetry for our 4D model, but it plays the role of time reversal symmetry for Dirac semimetals in 3D, as will be discussed in details in section \ref{sec:V}.

Up to now, we have identified a set of basic symmetries in our model,
\bea
{\cal S}=\{{\cal I},M_1,C_4,\Theta\}.
\label{Eq:Symmetry set S}
\eea
Next let us identify the minimal symmetry requirement to support this four-fold degeneracy. Our strategy is to add a perturbation term that breaks a certain symmetry $A$ in ${\cal S}$ (Eq. [\ref{Eq:Symmetry set S}]), and check the stability of SDP. If the SDP breaks into two Weyl points under $A$ symmetry breaking, we can identify $A$ as an essential symmetry to support the SDP, and vice versa. To start with, let us consider a generalized form of our TRI model,
\bea
H^{4D}=H^{4D}_0+H^{4D}_c+H^{4D}_k
\eea
where $H^{4D}_c$ is a collection of constant perturbation terms (leading order terms in ${\bf k}$), and $H^{4D}_k$ is the momentum dependent perturbations (higher order terms in ${\bf k}$). At this moment, we only require $H^{4D}_c$ and $H^{4D}_k$ to be $C_4$ invariant. For $H^{4D}_c$ terms, this requires $[H^{4D}_c,J_{12}]=0$. Based on Clifford algebra, only fifteen $\Gamma$ matrices satisfy this commutation relation.
Depending on whether a $\Gamma$ matrix commutes or anti-commutes with symmetries in $\{{\cal I},M_1,\Theta\}$, we can further classify them into eight different classes and check how they affect the stability of the SDP, as shown in Table. [\ref{tb:stability of surface DP}]. Stability of SDPs is checked by calculating the surface spectrum with the corresponding perturbation. In the stability column, a "$\checkmark$" indicates that a SDP is present, while a "$\times$" means that the SDP breaks into two Weyl fermions at different momentum positions.

\begin{table}[t]
\caption{Stability of surface Dirac point}
\begin{tabular}{c c c c c c c c}
\hline\hline \\
Class & Stability & $C_4$ & $M_1$ & ${\cal I}$ & $\Theta$ & ${\cal I}\Theta$ & $\Gamma$ matrices   \\[1ex]
\hline \\
$P_1$ & $\checkmark$ & $\checkmark$ & $\checkmark$ & $\times$ & $\times$ & $\checkmark$ & $\Gamma_3,\Gamma_{45}$  \\
$P_2$ & $\checkmark$ & $\checkmark$ & $\checkmark$ & $\times$ & $\checkmark$ & $\times$ & $\Gamma_4,\Gamma_{35}$  \\
$P_3$ & $\checkmark$ & $\checkmark$ & $\checkmark$ & $\checkmark$ & $\checkmark$ & $\checkmark$ & $\Gamma_5,\Gamma_{34},\Gamma_{345}$  \\
$P_4$ & N/A & $\checkmark$ & $\checkmark$ & $\checkmark$ & $\times$ & $\times$ & N/A \\
$P_5$ & $\times$ & $\checkmark$ & $\times$ & $\checkmark$ & $\times$  & $\times$ & $\Gamma_{12},\Gamma_{67},\Gamma_{125},\Gamma_{567}$ \\
$P_6$ & $\times$ & $\checkmark$ & $\times$ & $\times$ & $\checkmark$  & $\times$ & $\Gamma_{123},\Gamma_{367}$ \\
$P_7$ & $\checkmark$ & $\checkmark$ & $\times$ & $\times$ & $\times$ & $\checkmark$ & $\Gamma_{124},\Gamma_{467}$ \\
$P_8$ & N/A & $\checkmark$ & $\times$ & $\checkmark$ & $\checkmark$ & $\checkmark$ & N/A \\
\hline
\end{tabular}
\label{tb:stability of surface DP}
\end{table}

Due to the constraint of rotation symmetry, the higher order perturbation terms $H^{4D}_{k}$ has a general form
\bea
&&H^{4D}_{k}(k_{\pm},k_{3,4}) \nonumber \\
&=&\bp
0 & A(k_{3,4})k_+ & C(k_{3,4})k_-^2 & D(k_{3,4})k_- \\
A^{\dagger}(k_{3,4})k_- & 0 & E(k_{3,4})k_-^3 & F(k_{3,4})k_-^2 \\
C^{\dagger}(k_{3,4})k_+^2 & E^{\dagger}(k_{3,4})k_+^3 & 0 & B(k_{3,4})k_+ \\
D^{\dagger}(k_{3,4})k_+ & F^{\dagger}(k_{3,4})k_+2 & B^{\dagger}(k_{3,4})k_+ & 0 \\
\ep, \nonumber \\
&&
\label{Eq:H4D1}
\eea
where $A,B,C,D,E,F$ are all $2\times 2$ matrices whose detailed forms are to be determined, and will be constrained if additional symmetries are considered. We have denoted $k_{\pm}=k_1\pm ik_2$ for simplicity. Following the methods in Appendix B, $H^{4D}_{k}$ results in a perturbation term $\tilde{H}^{4D}_{k}$ in the surface Hamiltonian, which generally takes the form
\bea
&&\tilde{H}^{4D}_{k}(k_{\pm},k_{3}) \nonumber \\
&=&\bp
0 & \tilde{A}(k_{3})k_+ & \tilde{C}(k_{3})k_-^2 & \tilde{D}(k_{3})k_- \\
\tilde{A}^{\dagger}(k_{3})k_- & 0 & \tilde{E}(k_{3})k_-^3 & \tilde{F}(k_{3})k_-^2 \\
\tilde{C}^{\dagger}(k_{3})k_+^2 & \tilde{E}^{\dagger}(k_{3})k_+^3 & 0 & \tilde{B}(k_{3})k_+ \\
\tilde{D}^{\dagger}(k_{3})k_+ & \tilde{F}^{\dagger}(k_{3})k_+2 & \tilde{B}^{\dagger}(k_{3})k_+ & 0 \\
\ep. \nonumber \\
&&
\label{Eq:H4D1_surf}
\eea
Here $\tilde{A},\tilde{B},\tilde{C},\tilde{D},\tilde{E}$ are polynomial functions of $k_3$. Notice that $\tilde{H}^{4D}_{k}$ vanishes at $k_3$ axis, where $k_1=k_2=0$. Therefore, the appearance of $H^{4D}_{k}$ does not affect the stability of SDP at all. Together with the results in Table. [\ref{tb:stability of surface DP}], we conclude that a SDP is stable only when either in-plane mirror $M_1$ or the combined symmetry ${\cal I}\Theta_{3D}$ is present.
%However, our description seems to be not complete, as we can make the following observation: In Fig. \ref{Fig:4D dispersion} (a), it seems that two surface Weyl fermions living on the same surface are "always" coupled to form a surface Dirac point (SDP), as long as the following set of symmetry ${\cal S}=\{{\cal I},M_1,C_4,\Theta_{3D}\}$ are unbroken.
%Interestingly, a set of well-defined RCN only guarantees the gapless nature of surface states, but says nothing about the corresponding degeneracy properties. Therefore, this degeneracy (if not accidental) should be protected by symmetries other than $C_4$ rotation. %In this section, we identify this minimal symmetry requirements, and show that there are two types of SDPs stabilized by additional symmetries. These additional symmetries also put constraints to the values of RCNs.

%Every coin has two sides. The simple form of $H^{4D}_0$ is also a drawback, which leads to a very high symmetry of this model. As we have shown, $C_4$ is a necessary symmetry for the system, while the roles other symmetries are playing remain unclear. Generally, we could classify $C_4$ invariant perturbation terms $H^{4D}_1$ into two different types: momentum dependent terms $H^{4D}_{1,k}$ and momentum independent terms $H^{4D}_{1,C}$.

%Now let us study the effect of $H^{4D}_{1,c}$ with the help of Clifford algebra. The in-plane rotation generator is given by $J_{12}=\Gamma_{67}-\frac{1}{2}\Gamma_{12}$. In all 64 $\Gamma$ matrices, there are 15 of them that commutes with $J_{12}$. These 15 matrices exhaust possible forms of $H^{4D}_{1,c}$.

Alternatively, we can consider a generalization of Eq. [\ref{Eq:Block of H4D0}]:
\be
h_{\pm\frac{J}{2}}=
\bp
m_{\pm\frac{J}{2}}(k_3,k_4) & \pm v_{\pm\frac{J}{2}}(k_3-i k_4) \\
\pm v_{\pm\frac{J}{2}}(k_3+i k_4) & -m_{\pm\frac{J}{2}}(k_3,k_4) \\
\ep
\ee
for $J=1,3$. It is easy to show that either $M_1$ or ${\cal I}\Theta$ will put the following constraints:
\bea
m_{+\frac{J}{2}}(k_3,k_4)&=&m_{-\frac{J}{2}}(k_3,k_4) \nonumber \\
v_{+\frac{J}{2}}&=&v_{-\frac{J}{2}}.
\label{Eq:Symmetry constraints of H4D0}
\eea
Eq. [\ref{Eq:Symmetry constraints of H4D0}] guarantees the degeneracy of bulk states between $|\pm\frac{J}{2}\rag$ representations. One can perturbatively solve for the equations of surface states and confirm their degenerate nature. Eq. [\ref{Eq:Symmetry constraints of H4D0}] also leads to another important relation of RCN:
\be
C^{(1)}_{\frac{J}{2}}=C^{(1)}_{-\frac{J}{2}}
\label{Eq:symmetry on C1}
\ee
where we have taken advantage of the simple form of $H^{4D}_0$. A more general approach to prove Eq. [\ref{Eq:symmetry on C1}] is to apply symmetry constraints (either $M_1$ or ${\cal I}\Theta$) to the Berry curvature ${\cal F}_{\frac{J}{2}}(k_3,k_4)$ defined in the $k_3$-$k_4$ plane:
\bea
{\cal F}_{\frac{J}{2}}(k_3,k_4)={\cal F}_{-\frac{J}{2}}(k_3,k_4).
\eea
This immediately leads to Eq. \ref{Eq:symmetry on C1}.

Notice that TRS ${\cal T}$ is absent in our model. If ${\cal T}$ is present, it forces ${\cal F}_{\frac{J}{2}}(k_3,k_4)=-{\cal F}_{-\frac{J}{2}}(-k_3,-k_4)$, giving rise to a strong constraint that $C^{(1)}_{\frac{J}{2}}=-C^{(1)}_{-\frac{J}{2}}$. Together with Eq. [\ref{Eq:symmetry on C1}], we find that the existence of ${\cal T}$ yields a zero $C^{(1)}_{\pm\frac{J}{2}}$. Therefore, ${\cal T}$ is required to be broken to avoid trivial topology in our TRI model.

To conclude, based on $C_4$ rotation symmetry, we define the RCN to protect the gapless nature of chiral surface states on the $k_3$ axis, which live on the surface along the $x_4$ direction. This gives rise to a new type of 4D topological crystalline insulator protected by rotation symmetry with surface Weyl fermions. Two surface Weyl fermions with opposite chiralities can never annihilate without breaking $C_4$ symmetry. With only $C_4$, these Weyl fermions do not necessarily appear at the same momentum to form a SDP. Only when additional symmetries are included, SDPs are stabilized. We have demonstrated that two sets of inequivalent minimal symmetry requirements,
\bea
&\text{Type-I}:&\ \{C_4,{\cal I}\Theta\} \nonumber \\
&\text{Type-II}:&\ \{C_4,M_1\}.
\label{Eq:minimal symmetry}
\eea
can stabilize SDPs in our TRI model.

\section{Mapping a 4D TRI to a 3D DSM}\label{sec:V}

In this section, we will map our artificial 4D TRI system to a realistic 3D bulk DSM, following the construction scheme of 4D QHI.

First of all, we analytically solve for energy eigenfunction of surface states at $k_1=k_2=0$, and then project the bulk Hamiltonian onto the surface. Up to the first non-trivial order, we arrive at the effective surface Hamiltonian $H_{surf}^b$ ($H_{surf}^t$) for bottom (top) surface (See Appendix B for details),
\bea
H_{surf}^b&=&v_f\bp
k_3-k_0 & k_+ & 0 & 0 \\
k_- & -(k_3-k_0) & 0 & 0 \\
0 & 0 & -(k_3-k_0) & -k_+ \\
0 & 0 & -k_- & k_3-k_0 \\
\ep \nonumber \\
H_{surf}^t
&=&v_f\bp
k_3+k_0 & k_+ & 0 & 0 \\
k_- & -(k_3+k_0) & 0 & 0 \\
0 & 0 & -(k_3+k_0) & -k_+ \\
0 & 0 & -k_- & k_3+k_0 \\
\ep
\nonumber \\
&&
\label{Eq:Surface Hamiltonian}
\eea
on the basis function being $|\psi\rangle^{t(b)}=(|\frac{1}{2}\rag^{t(b)},|\frac{3}{2}\rag^{t(b)},|-\frac{3}{2}\rag^{t(b)},|-\frac{1}{2}\rag^{t(b)})^T$ for the top (bottom) surface.
\begin{figure}[t]
\centering
\includegraphics[width=0.5\textwidth]{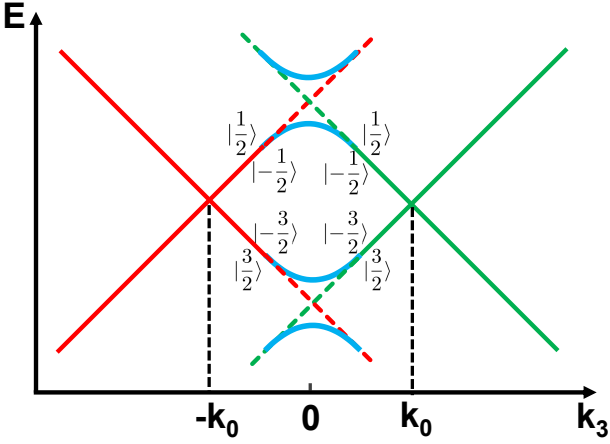}
\caption{In this plot, we demonstrate the mapping from 4D TRI thin film states to the 3D DSM state. Total angular momentum of SDP states are labelled next to the corresponding bands. Green (red) lines represent the SDP from top (bottom) surface. Hybridization effects between SDPs are plotted in blue lines.}
\label{Fig:3D dispersion}
\end{figure}

Now let us consider a thin film configuration along the $x_4$ direction, and SDPs from both surfaces will be projected together. $H_{shift}$ shift these SDPs along $k_3$ axis, and stabilize them via a momentum separation of $2k_0$. As shown in Fig. [\ref{Fig:3D dispersion}], the TRI thin film system now hosts two bulk Dirac points, which originate from both top surface SDP (green lines) and bottom surface SDP (red lines). Inter-surface hybridization generally exists and anti-crossing happens when two SDPs of the same representation cross at $k_3=0$ (dashed lines). In this case, the original surface degree of freedom now becomes the new valley degree of freedom in the 3D thin film. In this case, we find that Eq. [\ref{Eq:Surface Hamiltonian}] exactly reproduces the low-energy physics of rotational invariant 3D DSM systems which have been experimentally confirmed in Na$_3$Bi\cite{wang2012} and Cd$_3$As$_2$\cite{wang2013} systems.

Furthermore, it is easy to check that the following symmetries ${\cal S}=\{{\cal I},M_1,C_4,\Theta\}$ still holds when reducing the dimension of the system. By doing surface projections, we can explicitly reconstruct these symmetry operators in the surface bases,
\bea
&\tilde{{\cal I}}=&\tilde{s}_1\otimes\tilde{\sigma}_3\otimes\tilde{\tau}_3 \nonumber \\
&\tilde{M}_1=&i\tilde{s}_0\otimes\tilde{\sigma}_2\otimes\tilde{\tau}_1 \nonumber \\
&\tilde{C_4}=&\tilde{s}_0\otimes exp(i\tilde{J}_{12}\frac{\pi}{2})\nonumber \\
&\tilde{\Theta}=&i\tilde{s}_1\otimes\tilde{\sigma}_2\otimes\tilde{\tau}_1K
\label{Eq:3D Symmetry}
\eea
where we have defined a set of new Pauli matrices spanned in the surface bases. $\tilde{s}_i$, $\tilde{\sigma}_i$ and $\tilde{\tau}_i$ characterize valley, spin and orbital degree of freedom. The new rotation generator $\tilde{J}_{12}$ is defined as
\bea
\tilde{J}_{12}&=&\tilde{s}_0\otimes
\bp
\frac{1}{2} & 0 & 0 & 0 \\
0 & \frac{3}{2} & 0 & 0 \\
0 & 0 & -\frac{3}{2} & 0 \\
0 & 0 & 0 & -\frac{1}{2} \\
\ep
\eea
Now we find that the operator $\tilde{\Theta}$ is exactly the TRS operation in 3D DSM. Therefore, these symmetry operations (Eq. [\ref{Eq:3D Symmetry}]) are consistent with the symmetry operations in the effective theory of 3D DSMs.

Here we have assumed that the system possesses all the symmetry operations in ${\cal S}=\{{\cal I},M_1,C_4,\Theta\}$. In general, we can find a DSM state, whose corresponding 4D TRI is constructed by adding additional symmetry breaking terms to $H^{4D}_0$, as long as the minimal symmetry requirement (Eq. [\ref{Eq:minimal symmetry}]) is satisfied. For example, an inversion symmetry breaking term can be introduced by
\bea
H_{{\cal I}}^{4D}&=&D k_1(s_1\otimes\sigma_1-s_2\otimes\sigma_2)\otimes\tau_3 \nonumber \\
&&+D k_2(s_2\otimes\sigma_1+s_1\otimes\sigma_2)\otimes\tau_3.
\eea
In the effective surface Hamiltonian, $H_{{\cal I}}^{4D}$ produces a linear coupling between $|\frac{1}{2}\rag$ and $|-\frac{1}{2}\rag$, and the SDPs are still stable in this case. The new surface effective Hamiltonian is exactly mapped to the effective Hamiltonian of the inversion-symmetry-breaking phase of Cd$_3$As$_2$ (structure II phase)\cite{wang2013}.

In the earlier section \ref{sec:IV}, we have shown that there are two types of SDPs stabilized by different minimal symmetry requirements (Eq. [\ref{Eq:minimal symmetry}]). This automatically leads to two different mechanisms to protect 3D rotation-symmetric DSMs: (1) Type-I SDPs corresponds to type-I DSMs protected by $C_4$ and ${\cal I}\Theta$ symmetry. Examples are Na$_3$Bi\cite{wang2012} ($C_3$ symmetry instead of $C_4$ symmetry) and Cd$_3$As$_2$\cite{wang2013} (structure I). (2) Type-II SDP corresponds to type-II DSMs protected by $C_4$ and $M_1$ symmetry. An example of this type is the inversion-symmetry-breaking phase of Cd$_3$As$_2$\cite{wang2013} (structure II).

Most importantly, this mapping offers us a new approach to define the topological invariant of 3D DSMs. In earlier works, people were arguing about whether DSMs are topological or not, since conventional band topology is only well-defined for a gapped system. Various approaches have been tried to define this topological invariant, including classifying the topology of Fermi arc states\cite{yang2014}, as well as topological charges based on the rotational eigen-values of occupied bands\cite{yang2015b}. However, in both theories, only type-I DSMs are studied. By interpreting 3D DSMs as gapless surface states of a 4D gapped topological crystalline phases, the topological nature of DSMs is naturally revealed from the bulk-boundary correspondence. Topological invariants, RCNs, only relies on the existence of rotation symmetry, which naturally unifies the theory of two different types of DSMs.

\section{Conclusion and Discussions}\label{sec:VI}

In conclusion, we have constructed a model of a 4D TRI and map this system onto a 3D rotation-symmetry-protected DSM. A new 4D topological invariant, rotation Chern number, is defined in the presence of in-plane rotation symmetry, which well explains the topological nature of 3D DSMs. We have also identified the necessary conditions to realize SDPs in a 4D TRI. First of all, the RCN of the system must be nontrivial. Secondly, at least one of the minimal symmetry requirements (Eq. [\ref{Eq:minimal symmetry}]) must be satisfied. This immediately leads to two different protection mechanisms of 3D DSMs with rotation symmetries.

This idea of TRI can be easily generalized to a 4D system with other rotation symmetries $C_n$ ($n=2,3,6$). We notice that for any n-fold rotation symmetry, the state with angular momentum $\frac{J}{2}$ and that with angular momentum $\frac{J}{2}+n$ belong to the same representation. As a consequence, these two states can be coupled via a constant term without breaking $C_n$ symmetry in the effective model. In a $C_4$ or $C_6$ symmetric TRI system, one can always find a choice of bases that belong to different representations, and thus avoid this problem. However, for a $C_2$ symmetric TRI, such coupling terms generally exist and a well-defined SDP is absent. In a $C_3$ symmetric system, one can couple $|\frac{3}{2}\rag$ state and $|-\frac{3}{2}\rag$ state in a similar way. However, ${\cal I}\Theta$ can forbid such coupling terms, and stabilize only the type-I SDPs in this system. An example of the corresponding 3D DSM state is Na$_3$Bi. This concludes the classification of point-group-symmetric 4D TRIs protected by $C_n$ symmetry and the corresponding 3D DSMs. A future direction is to apply our proposed framework to systems with space group symmetries, which will lead us to the topological nature of DSMs protected by non-symmorphic symmetries\cite{young2012,steinberg2014}, as well as magnetic group symmetries \cite{tang2016}.

Although the definition of RCN relies on the extra auxiliary spatial dimension, it does have physical consequence on the theory of realistic 3D DSMs. For example, the effective action of 4D Dirac model is a 4D Chern-Simons term, from which the electro-magnetic (EM) response can be easily obtained\cite{qi2008}. This result is consistent with the EM response theory of 3D WSMs, giving rise to the famous chiral anomaly phenomenon\cite{zyuzin2012}. In analogous, one can also construct a similar effective theory for 4D TRIs. We expect the EM response of each representation follows that of a 2D QHI in the $x_3$-$x_4$ space, because of the RCN. Furthermore, additional symmetries (Eq. [\ref{Eq:minimal symmetry}]) will put constraints on the form of EM response.

On the other hand, K-matrix formulism has been proved to be successful in classifying 2D symmetry protected topological (SPT) phases\cite{lu2012}. Despite the fact that it is difficult to directly apply this formulism to DSMs in 3D, we can write down the bulk effective theory of a TRI with the K-matrix formulism in 4D. This naturally extend our discussion to an interacting TRI system, which corresponds to an interacting DSM system in 3D. Recently, similar interacting TCI systems have been shown to admit a classification reduction\cite{fidkowski2010,yao2013,qi2013,ryu2012}. For example, topological classification of a 2D topological mirror insulator will be reduced from $\mathbb{Z}$ to $\mathbb{Z}_4$ when interactions are incorporated\cite{isobe2015}. It is interesting to ask that how a similar classification reduction will happen in a 4D interacting TRI system. Consequently, one may wonder what appropriate interactions can be added to gap out a bulk DSM with an appropriate number of bulk Dirac points, without introducing any symmetry breaking orders and intrinsic topological orders. Our theory provide a completely new perspective to answer these interesting questions, which will be left for future works.

\bibliography{4DTCI}

\appendix

\section{General discussions on symmetry operations}

In this section, we will discuss the construction of symmetry operations for our 4D TRI model with the help of Clifford algebra. To start with, let us first briefly review how to construct symmetry operation for $4\times4$ $\gamma$ matrices in the effective model of 3D topological insulators (TI), such as Bi$_2$Se$_3$ \cite{liu2010}:
\be
H_{TI}=\sum_{i=1}^{3}v_i k_i\gamma_i+m(k)\gamma_5.
\ee
This is the effective model describing topological insulators in the Bi$_2$Se$_3$ family. Mirror symmetry $M_i$ and the in-plane rotation symmetry generator are given by
\bea
&M_i=&\gamma_i\gamma_4 \nonumber \\
&J_{12}=&-\frac{1}{2i}\gamma_1\gamma_2=-\frac{1}{2}\gamma_{12}.
\label{Eq:Generator_3DTI}
\eea
Here we make some remarks about a key difference between $H_{TI}$ and $H^{4D}_{Dirac}$, where the same set of $4\times 4$ matrices are used. In 4D Dirac model, momentum $k_4$ in the extra spatial dimension couples to $\gamma_4$, which constrains the form of mirror symmetry operation. As a result, 4D Dirac model will naturally break mirror symmetries which is defined in Eq. \ref{Eq:Generator_3DTI} of a 3D TI model. Interestingly, mirror symmetry $M_1$ can be restored by constructing the 4D TRI Hamiltonian with $8\times 8$ $\Gamma$ matrices, as will be shown later.

\subsection{Rotation Symmetry Generator}

Our TRI model is given by $H^{4D}_0=k_1\Gamma_1-k_2\Gamma_2+k_3\Gamma_3+k_4\Gamma_4+m(k)\Gamma_5$ in the continuum limit, where $\Gamma_i$ is an $8\times8$ $\Gamma$ matrix. Under the bases defined in the main text, it is easy to see the rotation generator $J_{12}$ in the $x_1$-$x_2$ plane is simply given by the $S_z$ spin operator for a spin-$\frac{3}{2}$ particle,
\bea
J_{12}&=&S_z \nonumber \\
&=&\bp
\frac{1}{2} & 0 & 0 & 0 \\
0 & \frac{3}{2} & 0 & 0 \\
0 & 0 & -\frac{3}{2} & 0 \\
0 & 0 & 0 & -\frac{1}{2} \\
\ep\otimes \tau_0 \nonumber \\
&=&\Gamma_{67}-\frac{1}{2}\Gamma_{12}.
\eea
Interestingly, the form of $J_{12}$ is slightly different from that of the 3D TI (Eq. \ref{Eq:Generator_3DTI}). In Bi$_2$Se$_3$, the bases are $(|+,\frac{1}{2}\rag, |+,-\frac{1}{2}\rag, |-,\frac{1}{2}\rag, |-,-\frac{1}{2}\rag)$, where "$\pm$" represents the parity of the state. Compared with the bases for 4D TRI, this difference in $J_{12}$ originates from different angular momentum of the bases states.

To understand this difference, let us start with the general rotation properties of $\Gamma$ matrices. We can first assume the rotation generator to be $J$, which is to be determined, and then rotate $\Gamma_i$ by an infinitesimal angle $\theta$ with the help of the rotation generator,
\be
\Gamma_i'(\theta)=e^{iJ\theta}\Gamma_i e^{-iJ\theta}=(1+iJ\theta)\Gamma_i (1-iJ\theta)=\Gamma_i +i\theta[J,\Gamma_i].
\ee
Reorganizing this expression to a differential equation, we arrive at
\be
\frac{d\Gamma_i'(\theta)}{d\theta}=i[J,\Gamma_i].
\label{Eq:rotation of Gamma}
\ee
In our model, the momentum $k_i$ is only coupled to $\Gamma_i$. For a in-plane rotation operation $R_{\theta}=e^{iJ_{12}\theta}$, it transforms $k_{1,2}$ to a linear combination of $k_1$ and $k_2$, while leaving $k_{3,4}$ invariant. This is represented by
\bea
\bp
k'_+ \\
k'_- \\
k'_3 \\
k'_4 \\
\ep
=
\bp
e^{i\theta} & 0 & 0 & 0 \\
0 & e^{-i\theta} & 0 & 0 \\
0 & 0 & 1 & 0 \\
0 & 0 & 0 & 1 \\
\ep
\bp
k_+ \\
k_- \\
k_3 \\
k_4 \\
\ep.
\eea
Therefore, to make $H^{4D}_0$ invariant, we have
\bea
\bp
\Gamma'_+ \\
\Gamma'_- \\
\Gamma'_3 \\
\Gamma'_4 \\
\ep
=
\bp
e^{-i\theta} & 0 & 0 & 0 \\
0 & e^{i\theta} & 0 & 0 \\
0 & 0 & 1 & 0 \\
0 & 0 & 0 & 1 \\
\ep
\bp
\Gamma_+ \\
\Gamma_- \\
\Gamma_3 \\
\Gamma_4 \\
\ep.
\eea
Compared with Eq. \ref{Eq:rotation of Gamma}, we arrive at
\begin{eqnarray}
[J_{12},\Gamma_{1}]=i\Gamma_{2},\ \ [J_{12},\Gamma_2]=-i\Gamma_{1}.
\end{eqnarray}
This implied that $J_{12}$ can be written as
\be
J_{12}=-\frac{1}{2}\Gamma_{12}+J_{12}^0.
\ee
where we require $[J_{12}^0,\Gamma_{1,2,3,4,5}]=0$. Here $J_{12}^0$ generally exists, while it is set to be zero for Bi$_2$Se$_3$ systems. A natural choice of $J_{12}^0$ is the product of $\Gamma_6$ and $\Gamma_7$. We define $J_{12}^0=c\times \Gamma_{67}$. Here $c$ is a constant that controls the angular momentum of basis functions. In this case, $J_{12}$ generates the rotation matrix for the following bases: $(|-\frac{1}{2}+c\rangle, |\frac{1}{2}+c\rangle, |-\frac{1}{2}-c\rangle, |\frac{1}{2}-c\rangle)$. Here we simply choose $c=1$ to match our bases choice.

\subsection{Mirror Symmetry}
Next, we will be focusing on the mirror symmetry $M_1$. Under $M_1$, the Hamiltonian $H^{4D}_0$ transform as $M_1H^{4D}_0({\bf k})M_1^{-1}=H^{4D}_0(-k_1,k_2,k_3,k_4)$. Therefore, we find that $M_1$ anti-commutes with $\Gamma_1$, while commuting with $\Gamma_{2,3,4,5}$. To satisfy this condition, $M_1$ has to take the form $M_1=\Gamma_1\Gamma_{6,7}$.

On the other hand, if we consider rotation operation in the $x$-$y$ plane $C_{\theta}$, we have $M_{1,2}C_{\theta}=C_{\theta}^{-1}M_{1,2}$. In terms of the generator $J_{12}$, we require that $\{J_{12},M_{1,2}\}=0$. It is easy to check that $M_1=\Gamma_1\Gamma_{6,7}$ satisfies this condition. In our discussion, we have chosen $M_{1}=\Gamma_{1}\Gamma_6$, which is consistent with the theory of 3D DSMs.

\subsection{Absence of Time Reversal Symmetry}

In section \ref{sec:IV}, we have shown that TRS ${\cal T}$ is absent in our TRI model to avoid trivial topology. In this section, we show that TRS $\Theta$ is absent in our TRI model from a different perspective. In the Clifford algebra, we define TRS ${\cal T}$ as an anti-unitary operator ${\cal T}=UK$ that satisfies
\bea
{\cal T}\Gamma_{1,2,3,4}{\cal T}^{-1}&=&-\Gamma_{1,2,3,4} \nonumber \\
{\cal T}\Gamma_{5}{\cal T}^{-1}&=&\Gamma_{5}.
\eea
Here $U$ is an $8\times8$ unitary matrix and $K$ is complex conjugate operation. We first act on the complex conjugate operation, only $\Gamma_{2,4}$ flip their signs. Therefore, we require that matrix $U$ anti-commutes with $\Gamma_{1,3}$ and meanwhile commutes with $\Gamma_{2,4,5}$. As a result, $U$ matrix candidates are $\Gamma_1\Gamma_3$ and $\Gamma_2\Gamma_4\Gamma_5$.

In addition, TRS should commute with crystalline symmetries. As a result, {\cal T} is required to anti-commutate with the rotation symmetry generator $J_{12}$. However, neither $\Gamma_1\Gamma_3K$ nor $\Gamma_2\Gamma_4\Gamma_5K$ anti-commute with $J_{12}$. Therefore, we conclude that our TRI model does not have TRS symmetry.

\section{Effective surface spectrum of a 4D TRI}
In this section, we are going to solve the zero energy solutions on the surfaces of 4D TRI model. Analytically, the wavefunctions of surface zero modes at $k_1=k_2=k_3=0$ can be obtained. And we will apply these wave-functions as our new surface bases to perturbatively obtain the effective Hamiltonian for both surfaces. To start with, we rewrite our Hamiltonian as
\bea
H^{4D}_0(k_1,k_2,k_3,k_4)&=&
\bp
H^u & 0 \\
0 & H^d \\
\ep + H_{shift} \nonumber \\
&=&
\bp
H^u & 0 \\
0 & -H^u \\
\ep + H_{shift} \nonumber \\
H^u(k_1,k_2,k_3,k_4)&=&\bp
h_{\frac{1}{2}} & 0 \\
0 & h_{\frac{3}{2}} \\
\ep \nonumber \\
h_{\frac{1}{2}}(0,0,0,k_4)&=&v_f k_4\tau_2+ m(k)\tau_3 \nonumber \\
h_{\frac{3}{2}}(0,0,0,k_4)&=&v_f k_4\tau_2- m(k)\tau_3,
\eea
where $v_f$ is the isotropic Fermi velocity. The above block-diagonal form of $H(k)$ indicates that we can solve the zero mode for each $2\times2$ block independently. So we can take the following trial solution with an exponential spatial part and a two-component spinor part $\xi$. For example, $\psi_{\frac{1}{2}}$ for $h_{\frac{1}{2}}$ is
\be
\psi_{\frac{1}{2}}=e^{\lambda x_4}\xi.
\ee
For zero modes, we have
\bea
0&=&h_{\frac{1}{2}}\psi_{\frac{1}{2}} \nonumber \\
&=&[-iv_f \partial_4\tau_2+ (m_0+4B_0+\frac{B_0}{2}\partial_4^2)\tau_3]e^{\lambda x_4}\xi \nonumber \\
&=&[-iv_f \lambda\tau_2+(m_0+4B_0+\frac{B_0}{2}\lambda^2)\tau_3]\xi \nonumber \\
&=&[-v_f \lambda \tau_1+(m_0+4B_0+\frac{B_0}{2}\lambda^2)]\xi  \nonumber \\
&=&[-s v_f \lambda+(m_0+4B_0+\frac{B_0}{2}\lambda^2)]\xi_s,  \nonumber \\
\eea
where in the third line, we have multiply both sides by $\tau_3$. We have also defined $\tau_1\xi_s=s\xi_s$, with $s=\pm1$. Here, $\lambda$ can be solved as
\bea
&\lambda_{s,\pm}&=\frac{1}{B_0}(s v_f\pm\sqrt{v_f^2-2B_0 (m_0+4B_0)}) \nonumber \\
&\lambda_{s,+}+\lambda_{s,-}&= \frac{2sv_f}{B_0} \nonumber \\
&\lambda_{s,+}\lambda_{s,-}& = \frac{2(m_0+4B_0)}{B_0}.
\eea
A general solution would take the following form
\be
\psi_{\frac{1}{2}}(x_4)=\sum_{s}(c_{s,+}e^{\lambda_{s,+}x_4}+c_{s,-}e^{\lambda_{s,-}x_4})\xi_s.
\ee
We are ready to place boundary conditions to restrict the expression of $\psi_{\frac{1}{2}}(x_4)$. First, we require the wave function vanishes at the boundary ($x_4=0$), leading to $0=\psi_{\frac{1}{2}}(x_4=0)=\sum_{s}(c_{s,+}+c_{s,-})\xi_s$. As a result $c_{s,+}=-c_{s,-}$, and
\be
\psi_{\frac{1}{2}}(x_4)=\sum_{s}c_{s,+}(e^{\lambda_{s,+}x_4}-e^{\lambda_{s,-}x_4})\xi_s.
\ee
For the bottom surface, we choose $x_4 > 0$, and require surface state wave function vanishes as $x_4\rightarrow +\infty$. This requires $\lambda_{s,+}<0,\lambda_{s,-}<0$, leading to
\bea
s&=&-1,\ \ m_0(m_0+4B_0)>0 \nonumber \\
\psi_{\frac{1}{2}}^b&\propto& \xi_{-1} = (1,-1,0,0)^T.
\eea
Here the upper index $b$ denotes that $\psi_{\frac{1}{2}}^b$ is the wave-function for bottom surface. Similarly, for $h_{\frac{3}{2}}$, we simply switch the sign of fermi velocity $v_f\Rightarrow -v_f$. Thus the only difference is that we take $s=+1$, and $\psi_{\frac{3}{2}}^b\propto \xi_{+1} = (0,0,1,1)^T$. The effective Hamiltonian of bottom surface state is given by

\begin{widetext}
\bea
H_{surf}^b =\bp
\langle \psi_{\frac{1}{2}}^b|H^u|\psi_{\frac{1}{2}}^b\rangle & \langle \psi_{\frac{1}{2}}^b|H^u|\psi_{\frac{3}{2}}^b\rangle & 0 & 0 \\
\langle \psi_{\frac{3}{2}}^b|H^u|\psi_{\frac{1}{2}}^b\rangle & \langle \psi_{\frac{3}{2}}^b|H^u|\psi_{\frac{3}{2}}^b\rangle & 0 & 0 \\
0 & 0 & \langle \psi_{\frac{1}{2}}^b|H^d|\psi_{\frac{1}{2}}^b\rangle & \langle \psi_{\frac{1}{2}}^b|H^d|\psi_{\frac{3}{2}}^b\rangle \\
0 & 0 & \langle \psi_{\frac{3}{2}}^b|H^d|\psi_{\frac{1}{2}}^b\rangle & \langle \psi_{\frac{3}{2}}^b|H^d|\psi_{\frac{3}{2}}^b\rangle \\
\ep
=v_f \bp
k_3-k_0 & k_+ & 0 & 0 \\
k_- & -(k_3-k_0) & 0 & 0 \\
0 & 0 & -(k_3-k_0) & -k_+ \\
0 & 0 & -k_- & k_3-k_0 \\
\ep. \nonumber \\
\label{Eq:Bottom Surface}
\eea
\end{widetext}

For top surface, $x_4 < 0$, we require surface state wave function vanishes as $x_4\rightarrow -\infty$. This requires $\lambda_{s,+}>0,\lambda_{s,-}>0$, and
\bea
s&=&+1,\ \ m_0(m_0+4B_0)>0 \nonumber \\
\psi_{\frac{1}{2}}^t&\propto& \xi_{+1} = (1,1,0,0)^T.
\eea

For $h_{\frac{3}{2}}$, we take $s=-1$, and $\psi_{\frac{3}{2}}^t\propto \xi_{-1} = (0,0,1,-1)^T$. The effective Hamiltonian of the top surface state is given by
\bea
&&H_{surf}^t \nonumber \\
&=& v_f \bp
k_3+k_0 & k_+ & 0 & 0 \\
k_- & -(k_3+k_0) & 0 & 0 \\
0 & 0 & -(k_3+k_0) & -k_+ \\
0 & 0 & -k_- & k_3+k_0 \\
\ep. \nonumber \\
&&
\label{Eq:Top Surface}
\eea
$H_{surf}^t$ ($H_{surf}^b$) describes a massless 3D Dirac fermion at $k_3=k_0 (-k_0)$. In the thin film limit, the low energy bulk Hamiltonian near the Fermi level is given by
\bea
H_{eff}=\bp
H_{surf}^t & 0 \\
0 & H_{surf}^b \\
\ep.
\eea
which is exactly the effective Hamiltonian of 3D DSMs, such as Na$_3$Bi and Cd$_3$As$_2$.

\end{document}